# Fusion d'images : application au contrôle de la distribution des biopsies prostatiques

Pierre Mozer [1 et 2], Michaël Baumann [2 et 3], Grégoire Chevreau [1 et 2], Jocelyne Troccaz [2]
[1] Service d'urologie, Groupe hospitalier Pitié-Salpêtrière, Paris.
[2] Laboratoire TIMC/IMAG, Université Joseph-Fourier, La Tronche, Grenoble.
[3] Koelis, La Tronche, Grenoble.

Correspondant : Pierre Mozer, Service d'Urologie, CHU La Pitié-Salpêtrière, Bvd de l'Hopital, 75013 Paris.

E-mail : pierre.mozer@sls.aphp.fr

# INTRODUCTION

Les biopsies de prostate, réalisées le plus souvent par voie endorectale sous guidage échographique 2D, restent le seul moyen de confirmer le diagnostic de cancer et d'évaluer son pronostic.

L'objectif est de réaliser un échantillonnage probabiliste de la prostate sachant néanmoins que certaines équipes tentent actuellement, afin d'en améliorer la sensibilité, de diriger ces biopsies vers une cible déterminée à partir d'image IRM [1].

L'évaluation de la qualité de l'échantillonnage est donc un critère particulièrement important à prendre en compte pour la prise en charge des patients. Cette évaluation ne peut pas être réalisée en étudiant uniquement la position externe de la sonde d'échographie à cause des mouvements du patient, des mouvements de la prostate et de sa déformation sous l'action de la sonde endorectale.

Dans le cadre de cette évaluation, nous présentons les résultats concernant les capacités d'un opérateur à réaliser un planning de biopsies guidées par échographie endorectale 2D en nous basant sur la fusion d'images échographiques 3D.

# LA FUSION D'IMAGES

La fusion d'images consiste à mettre en correspondance des images d'une même modalité d'imagerie (fusion monomodale) acquises à des instants différents ou bien des images de modalités d'imagerie différentes (fusion multimodale).

Elle peut être rigide si l'on part du principe qu'il n'y a pas de déformation entre les différentes modalités mises en correspondance ou élastique dans le cas contraire.

- La fusion rigide consiste à rechercher 3 rotations et 3 translations permettant de superposer au mieux les images à fusionner. Elle peut-être recherchée entre une image volumique et une série d'images projectives 2D, par exemple pour superposer une image scanner et des images de fluoroscopie, ou entre deux images volumiques. Le résultat peut être exprimé sous la forme d'une matrice dite de transformation.

- La fusion élastique est plus complexe que la fusion rigide, car elle prend en compte des déformations globales et locales entre les modalités que l'on souhaite mettre en correspondance. Ce type de fusion peut être utilisé pour comparer des images de patients différents, mais aussi pour fusionner différentes modalités entre elles afin de s'affranchir des différentes déformations et distorsions liées aux capteurs. Ce type de fusion permet également de suivre des déformations d'organes avec le temps. Le résultat peut être exprimé sous la forme d'un champ de vecteurs de déplacement plus ou moins dense.

Qu'elle soit rigide ou élastique, la fusion peut être réalisée à partir de points sélectionnés, dans l'image ou directement à partir de l'analyse des pixels constituant l'image.

- La première méthode de fusion consiste à sélectionner des points dans les deux images à fusionner. Le clinicien peut sélectionner des points, en général 3 ou 4, situés à des endroits jugés comme identiques dans les deux images. Les points localisés dans les deux images forment alors des couples de points homologues et pour fusionner les images on calcule la transformation minimisant la distance entre les couples de points homologues. On peut aussi sélectionner dans chaque image un nombre de points différents – par exemple délimitant le pourtour d'un organe, la sélection de ces points étant appelée segmentation. On obtient alors deux nuages de points et on calcule la transformation superposant au mieux ces nuages de points en minimisant leur distance. Pour que le calcul de cette distance soit pertinent, il est souvent nécessaire de sélectionner un nombre important de points qui varie en fonction de la forme de l'organe sur lequel la fusion est appliquée. En général, au moins une centaine de points sont nécessaires. Pour le clinicien, il s'agit d'une tâche particulièrement longue si elle est réalisée manuellement. Elle peut parfois être automatisée, mais les algorithmes doivent être capables de s'adapter aux variations anatomiques, quelles soient congénitales ou acquises, à la réponse spécifique du patient (comme l'échogénéïcité des tissus) ainsi qu'aux paramètres d'acquisition des images. Il n'est pas rare que les méthodes de segmentation entièrement automatiques soient mises en défaut.

Dans le cadre de l'appariement de points homologues, le choix de ceux-ci influe de façon notable sur la précision de la fusion. Dans le cadre de la fusion par la mise en correspondance de nuages de points, la méthode est plus précise, mais le temps nécessaire à la segmentation rend cette étape souvent difficilement réalisable en routine clinique.

- La seconde méthode consiste à appliquer ce que l'on appelle une fusion iconique. C'est une approche qui ne nécessite pas d'intervention humaine et qui repose sur la mesure de la similarité entre les deux images. Cette mesure de similarité est une équation mathématique qui est fonction des modalités d'images à fusionner et qui cherche à modéliser la relation entre le niveau de gris d'une image et le niveau de gris dans l'autre image.

Il est important de comprendre que la fusion est appliquée à une image acquise à un temps donné. Ainsi par exemple, en cas de fusion avec des images échographiques, les mouvements des différents organes ne peuvent être pris en compte en temps réel que si la méthode de mise en correspondance, qu'elle soit iconique ou réalisée à l'aide de points, est appliquée pour chacune des images obtenues en temps réel par l'appareil d'échographie (on peut estimer que les images sont obtenues à une fréquence de 30 Hz). Cela rend impossible la fusion temps réel basé sur la mise en correspondance utilisant des points, puisque la détermination de ces points nécessite le plus souvent une intervention humaine. Les temps de calcul actuels des méthodes iconiques ne permettent pas encore d'avoir des systèmes temps réel. Les systèmes de fusion actuels permettent donc uniquement d'obtenir une fusion précise soit *a posteriori*, soit à un temps donné particulier en partant du principe que les mouvements des organes sont reproductibles, comme c'est par exemple le cas pour le rein avec la respiration.

Dans le cadre de notre évaluation de la localisation *a posteriori* des trajets de biopsies au sein d'une image échographique 3D, nous employons pour fusionner les images une approche iconique. Cette technique est donc entièrement automatique ; pour l'instant elle ne prend pas en compte d'éventuelles déformations.

**MÉTHODE**

Le protocole consiste, au cours d'une procédure standard de 12 biopsies prostatique guidées en 2D, à réaliser une acquisition échographique volumique de la prostate après chaque biopsie afin de disposer de 12 volumes que nous fusionnons dans un volume de référence R0 acquis en début de procédure (cf. *fig.1*). La fusion de ces 12 volumes dans un volume de référence nous permet ensuite de comparer les biopsies réalisées par rapport aux biopsies planifiées.

**Patients**

Après accord du comité d'éthique, 15 patients devant bénéficier de biopsies de prostate, suivant les recommandations de l'association française d'urologie, ont été inclus dans cette étude. Les critères d'inclusion étaient un taux de PSA supérieur à 4 ng/ml chez des patients de moins de 75 ans pouvant bénéficier d'une prise en charge par un traitement curateur.
Les biopsies ont été précédées deux heures avant d'une antibioprophylaxie par une fluoroquinolone de type ciprofloxacine.

**Appareil d'échographie**

Nous avons utilisé un appareil Voluson-i (General Electric) avec une sonde endorectale RIC 5-9. La sonde RIC 5-9 est une sonde endorectale de type « endfire », c'est-à-dire que le transducteur est situé au bout de la sonde. La fréquence de la sonde est variable entre 5 et 9 MHz. Elle est constituée par une matrice linéaire de transducteurs qui tourne mécaniquement. Ce type d'appareil permet d'obtenir un volume échographique complet en 5 secondes.

**Protocole d'acquisition**

Un seul opérateur a réalisé l'ensemble des biopsies. Elles ont été effectuées en position gynécologique suivant le schéma habituel de 12 biopsies après une anesthésie locale comprenant l'injection de 10 cc de xylocaïne 1 % au contact des bandelettes vasculeux-nerveuses droite et gauche. Par rapport à une série de biopsies standard, le protocole nécessitait en plus : 1) en début de procédure, une acquisition d'un volume 3D de référence dit R0. C'est dans ce volume qu'est fusionné l'ensemble des biopsies réalisées ; 2) une acquisition d'un volume 3D après chaque biopsie pour visualiser le trajet de la biopsie.

Le fait d'effectuer l'acquisition volumique aiguille en place dans la prostate, en plus de permettre une visualisation très nette de l'aiguille, diminue fortement les mouvements de la prostate.

Pour chaque acquisition volumique, afin de diminuer autant que possible les problèmes liés aux mouvements, l'opérateur appuyait la sonde sur le bord de la table et demandait au patient de ne pas respirer. Un soin tout particulier a été apporté à ne pas comprimer la prostate avec la sonde de façon à limiter au maximum les déformations.

L'acquisition du volume de référence était faite comme si l'opérateur souhaitait réaliser des biopsies du lobe droit. Pour la réalisation des biopsies du lobe gauche, la sonde était tournée de 180° autour de son axe.

**Analyse des résultats**

Après fusion des 12 volumes échographiques acquis au cours des biopsies, le volume échographique prostatique de référence R0 contenant l'ensemble des biopsies (cf. *fig.2*) a été reformaté dans le plan frontal au sein d'une boîte englobante limitée au volume de la prostate. Cette boîte englobante a elle-même été subdivisée en 12 boîtes de même taille permettant ainsi de déterminer au sein de l'image les cibles correspondant au planning (cf. *fig.3*). Pour chaque biopsie, la longueur de l'aiguille à l'intérieur de chaque cible a été mesurée afin de comparer le planning au geste réalisé.

# RÉSULTATS

**Profil clinique des patients**

Sur 15 patients, 6 cancers ont été diagnostiqués (40%). L'âge, le taux de PSA et le volume prostatique sont reportés dans le *tableau I*.

Aucun effet secondaire des biopsies n'a été observé. La douleur à été rapportée comme étant en moyenne de 1,5 sur une échelle visuelle analogique côté de 0 à 10.

**Robustesse et précision de la fusion**

L'ensemble des volumes échographiques a été analysé. Chaque recalage a pris en moyenne 6 secondes sur un ordinateur standard et il a fonctionné dans plus de 96 % des cas. Les échecs de fusion étaient dus à des images échographiques de mauvaise qualité.

La précision de la fusion des images a été calculée en mesurant la distance entre des calcifications nettement individualisables dans deux images fusionnées. La précision mesurée sur l'ensemble des fusions a été notée comme étant en moyenne de 1,41 mm (max=3,84 mm). Dans 10 cas, il a été possible de visualiser le trajet d'une même biopsie dans deux volumes échographiques différents. La fusion des volumes et la comparaison des trajectoires ont mis en évidence une erreur moyenne de 4° et maximum de 10°.

**Résultats des biopsies**

En moyenne la cible a été atteinte dans 63 % des cas. Le taux de succès atteint 100 % pour la cible moyenne et parasagitale et il est plus faible pour les cibles latérales que pour les cibles parasagitales (cf. *tableau II*). Le taux de succès particulièrement faible pour la cible située à l'apex latéral peut-être expliqué par le fait que celle-ci contient peu de prostate.

Le faible pourcentage de longueur de biopsie à l'intérieur des cibles tend à prouver une faible concordance entre un planning de biopsies qui sont théoriquement parallèles et leur réalisation qui est contrainte par l'accès transrectal.

**CONCLUSION**

Cette étude permet de mettre en évidence qu'il existe une imprécision notable entre le planning prévu et les biopsies réellement réalisées. Ainsi, si le manque de précision des biopsies de prostate a déjà été mis en évidence par la faible corrélation avec les pièces de prostatectomies radicales [2], les mesures données par la méthode de recalage mise en œuvre sont plus précises. Une possibilité potentiellement intéressante de cette technique est de pouvoir fusionner différentes séries de biopsies pour s'assurer que des zones différentes ont bien été prélevées.

Qui plus est, ces données peuvent être exploitées pour vérifier par exemple *a posteriori* qu'une zone douteuse en IRM a bien été ponctionnée en effectuant une fusion multimodale entre le volume de référence R0 contenant les biopsies et une image volumique IRM. Cette technologie est actuellement maîtrisée en employant une fusion élastique par nuages de points [3] et fait l'objet d'une étude clinique prospective.

La véritable rupture technologique viendra lorsqu'il sera possible d'acquérir et de fusionner en temps réel des volumes échographiques permettant ainsi de suivre au sein d'une image échographiques 3D une cible au sein d'une prostate qui bouge et qui se déforme. Les progrès réalisés en quelques années dans le domaine de la fusion d'images permettent d'espérer à moyen terme de disposer d'un tel outil, qui jouera probablement un rôle important dans la prise en charge du cancer de la prostate, que ce soit dans un but diagnostic, mais aussi thérapeutique et améliorant la précision des traitements percutanés.

**Tableau I :** Âge, PSA et volume prostatique.

|  | Moyenne | Minimum | Maximum |
|---|---|---|---|
| Âge (années) | 66 | 66 | 75 |
| PSA (ng/ml) | 7,4 | 4 | 17,8 |
| Volume prostate (cc) | 39 | 20 | 115 |

**Tableau II : Résultats des zones biopsiées par rapport au planning.**

| Cibles | Nombre de biopsies vers la cible | Nombre de biopsies à l'intérieur de la cible | % moyen de biopsie à l'intérieur de la cible | Longueur moyenne de biopsie à l'intérieur de la cible (mm) | % moyen de la longueur de la biopsie à l'intérieur de la cible |
|---|---|---|---|---|---|
| BL | 29 | 16 | 55 % | 13 | 61 % |
| BP | 28 | 17 | 61 % | 14 | 62 % |
| ML | 29 | 23 | 79 % | 14 | 64 % |
| MP | 29 | 29 | 100 % | 16 | 71 % |
| AL | 29 | 9 | 31 % | 7 | 33 % |
| AP | 28 | 14 | 50 % | 13 | 61 % |
| Total ou moyenne | 172 | 108 | 63 % | 14 | 62 % |

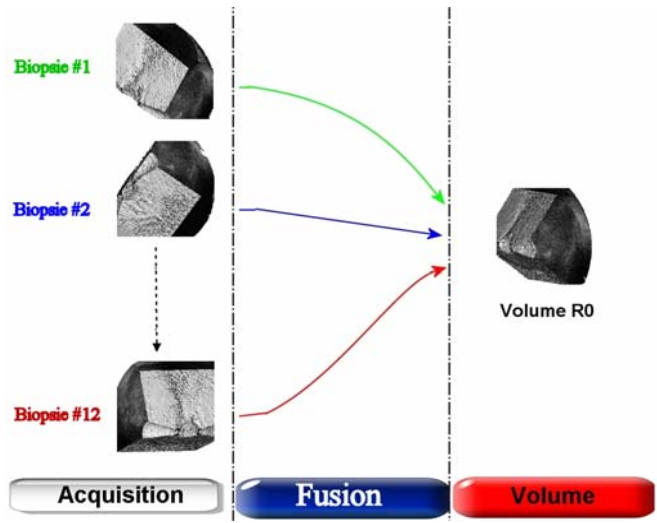

Fig. 1 : Fusion des 12 volumes échographiques acquis lors dès biopsies dans un seul volume échographique de référence.

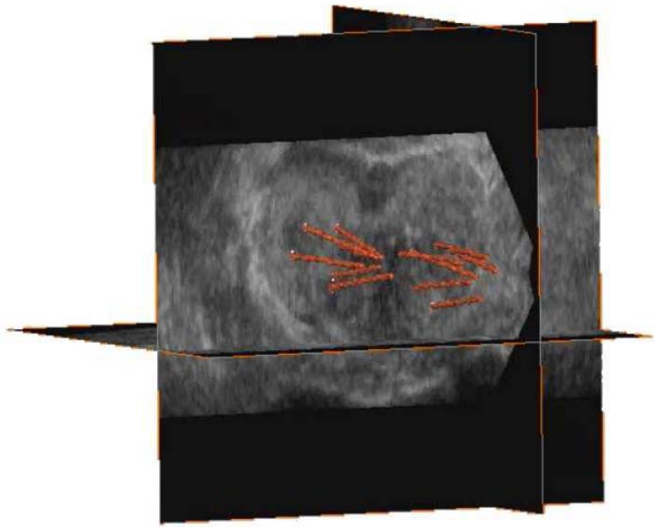

Fig. 2 : Projection des biopsies dans le volume échographique de référence R0.

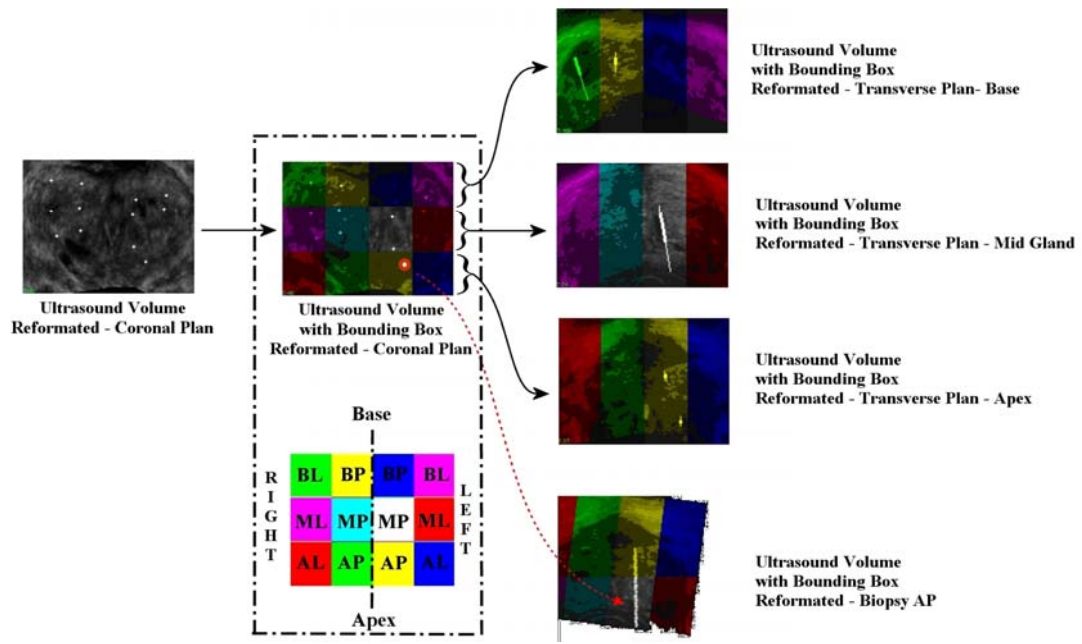

Fig. 3 : Création des cibles par rapport au planning des biopsies.